\documentclass{aa}
\usepackage{epsfig}
\usepackage{psfig}
\usepackage{graphicx}
\usepackage{txfonts}

\usepackage{lscape}

\titlerunning{An X-ray trail associated with PSR J2124$-$3358}
\newcommand{\PSR}{PSR~J2124$-$3358}

\begin{document}
\title{Searches for diffuse X-ray emission around millisecond pulsars: 
 An X-ray nebula associated with \PSR}

\author{C. Y. Hui \and W. Becker}   
\date{Received 21 November 2005 / Accepted 19 January 2006}
\institute{Max-Planck Institut f\"ur Extraterrestrische Physik, 
          Giessenbachstrasse 1, 85741 Garching bei M\"unchen, Germany}

\abstract{We report on diffuse X-ray emission associated with the 
 nearby solitary millisecond pulsar \PSR\ detected with XMM-Newton and Chandra. 
 The emission extends from the pulsar to the northwest by $\sim 0.5$ arcmin. 
 The spectrum of the nebular emission can be modeled with a  power-law of 
 photon index $2.2\pm0.4$, in line with the  emission originating from 
 accelerated particles in the post shock flow. For PSR J0437$-$4715, 
 PSR J0030+0451 and PSR J1024$-$0719, which all have spin parameters 
 comparable to that of \PSR, no diffuse emission is detected down 
 to a $3-\sigma$ limiting flux of $\sim 4-7\times10^{-15}$ ergs s$^{-1}$ cm$^{-2}$.
 
 }

\maketitle

\section{Introduction}

 Diffuse plerionic emission is often a special marker of young and powerful pulsars 
 if their spin-down energy is in excess of $\sim 10^{36}$ erg/s. Extended diffuse
 emission associated with older and less powerful pulsars, however, is rare and so 
 far only seen in Geminga (Caraveo et al.~2003), PSR B1929+10 (Becker et al.~2005), 
 PSR B2224+65 (Romani et al.~1997; Chatterjee \& Cordes 2002) and the millisecond 
 pulsars (MSPs) PSR B1957+20 (Stappers et al.~2003), J0437-4715 (Bell et al.~1995),  
 and J2124$-$3358 (Gaensler, Jones \&  Stappers 2002).
 While all three MSPs have bright bow-shock nebulae detected in  $H_\alpha$, diffuse 
 X-ray emission associated with it could be detected only from the black widow pulsar
 PSR B1957+20 in a recent Chandra observation (Stappers et al.~2003).  Comparable deep 
 observations of almost all X-ray bright MSPs, however,  have been performed by 
 XMM-Newton in previous years. These XMM-Newton data thus provides us a valuable 
 data base to search for diffuse and extended emission components from these MSPs
 as well. 

 In this letter we report on a search for diffuse X-ray emission to be associated with 
 the MSPs \PSR, J0437-4715, J0030+0451 and J1024-0719. All these pulsars have comparable 
 spin parameters (cf.~Table 1) so that differences in their emission properties are most 
 likely cause by differences in the pulsar-ISM interaction/local environment rather than by 
 differences in their total energy output. 

 PSR J2124-3358 and J1024-0719 were both discovered during the Parkes 436 MHz survey of 
 the southern sky (Bailes et al.~1997). PSR J0030$+$0451 was discovered at 430 MHz 
 during the Arecibo Drift Scan Search (Somer 2000) and independently in the Bologna 
 sub-millisecond pulsar survey (D'Amico 2000), whereas PSR J0437-4715 was discovered in 
 the Parkes southern sky survey (Johnston et al.~1993). PSR J0437-4715 was the first 
 MSP of which pulsed X-ray emission was detected (Becker \& Tr\"{u}mper 1993). X-ray 
 emission  from \PSR\ and J1024-0719 was reported by Becker \& Tr\"{u}mper (1998, 1999) 
 in ROSAT HRI data whereas the X-ray counterpart of PSR J0030$+$0451 was discovered in 
 the final ROSAT PSPC observation (Becker et al.~2000). A $H_\alpha$ bow shock is seen 
 around PSR J0437-4715 (Bell et al.~1995). However, its X-ray counterpart was not detected 
 by ROSAT and Chandra (Becker \& Tr\"umper 1999, Zavlin et al.~2002). Gaensler, Jones 
 \&  Stappers (2002) discovered an $H_\alpha$-emitting bow shock nebula around 
 PSR J2124-3358. This bow shock is very broad  and highly asymmetric about the
 direction of the pulsar's proper motion. The asymmetric shape might be caused by a 
 significant density gradient in the ISM, bulk flow of ambient gas and/or anisotropies 
 in the pulsar's relativistic wind (Gaensler, Jones \&  Stappers 2002). Observation of 
 PSR J0030+0451, performed with the ESO NTT in La Sila, did not detect diffuse $H_\alpha$ 
 emission associated with it (A.~Pellizzoni priv.~com.). 

 \begin{table*}
 \centering
 \caption{Pulsar Parameters of PSRs J0030+0451, J2124-3358, J1024-0719 and J0437-4715 (from Manchester et al.~2005) \label{ephemeris}}
 \begin{tabular}{lcccc}
 \hline\hline
 Pulsar                       & PSR J0030+0451 & PSR J2124-3358 & PSR J1024-0719 & PSR J0437-4715\\
 \hline
 Right Ascension (J2000)  & $00^{\rm h} 30^{\rm m} 27.432^{\rm s}$ & $21^{\rm h} 24^{\rm m} 43.862^{\rm s}$ & $10^{\rm h} 24^{\rm m} 38.700^{\rm s}$ & $04^{\rm h} 37^{\rm m} 15.787^{\rm s}$\\
 Declination (J2000)          & $+04^\circ\; 51'\; 39.7"$ & $-33^\circ\; 58'\; 44.257"$ & $-07^\circ\; 19'\; 18.915"$ & $-47^\circ\; 15'\; 08.462"$\\
 Pulsar Period, $P$ (ms)        & 4.865453207369  & 4.9311148591481 & 5.1622045539088 & 5.7574518310720  \\
 Period derivative $\dot{P}$ ($10^{-20}$ s s$^{-1}$) & 1.0 & 1.33 & 1.85 & 1.87 \\
 Age ($10^{9}$ yrs)          & 7.71 & 5.86 & 4.41 & 4.89\\
 Surface dipole magnetic field ($10^{8}$ G) & 2.23 & 2.60 & 3.13 & 3.32 \\
 Epoch of Period (MJD)      &  50984.4 & 50288.0 & 51018.0 & 51194.0 \\
 Dispersion Measure (pc cm$^{-3}$)  & 4.3328  & 4.6152 & 6.491 & 2.6469 \\
 Dispersion based distance (pc)   &  230 & 250 & 350 & 140            \\
 Spin-down Luminosity ($10^{33}$) ergs s$^{-2}$ & 3.43 & 4.38 & 5.3 & 3.87\\
 \hline
 \end{tabular}
 \end{table*}

\section{Observations and data analysis}
 The pulsars PSR J2124-3358, J0030+0451, J1024-0719 and J0437-4715 were observed by XMM-Newton 
 in 2002 April 14-15, in 2001 June 19-20, in  2003 December 2 and in 2002 October 9, respectively. 
 In all observations the MOS1/2 cameras were operated  in full-frame mode while the PN camera was 
 setup to work in the fast-timing mode. The PN fast-timing mode provides only limited spatial 
 information so that these data are not used in the present analysis. To block stray light and 
 optical leakage from bright foreground stars we operated the MOS1/2 cameras with the thin 
 filters during the observations of PSRs J0030+0451, J1024-0719 and J0437-4715 while 
 PSR J2124-3358 was observed by using a medium filter. Events were selected for the energy 
 range $0.25-10$ keV and standard correction procedures were applied to reduce the data 
 (e.g.~Becker et al.~2005). Filtering the data for times of excessive background from soft 
 proton flares was done using conservative thresholds in view of the planned search for 
 diffuse and extended X-ray emission. The effective exposure time after data reduction 
 turns out to be 40 ksec for \PSR, 14 ksec for J0030+0451, 65 ksec for J1024-0719 and 
 55 ksec for J0437-4715.

 A vignetting corrected image of the field of PSR J2124-3358 as seen by the XMM-Newton's 
 MOS1/2 CCDs is shown in Figure 1a. The binning factor in this image is 6 arcsec. 
 Adaptive smoothing with a Gaussian kernel of $\sigma<4$ pixels has been applied to
 the image. X-ray contours are calculated and overlaid on the image. The contour lines 
 are at the levels of $(4.2, 5.2, 7.6, 13, 28, 63)\times10^{-6}$ cts s$^{-1}$ arcsec$^{-2}$.  
 It can be seen that the X-ray source which is coincident with the pulsar position 
 has an asymmetric source structure of $\sim 0.5$ arcmin extent, with its orientation
 to the northwest. Systematic effects which could cause an adequate distortion of 
 the instrument's point spread function (PSF) are not known for XMM-Newton. We are 
 therefore prompted to interpret this elongated structure in terms of a pulsar X-ray 
 trail. The net count rate of the diffuse emission in a $30\times 35$ arcsec box 
 centered at RA(J2000)=$21^{\rm h} 24^{\rm m} 42.69^{\rm s}$, Dec=$-33^\circ 58' 16.06"$ and oriented 
 along the extended feature is estimated to be  $(1.25\pm0.16)\times10^{-3}$ cts s$^{-1}$. 
 For this estimate we determined the background from a low count region close to \PSR.
 For comparison, the pulsar emission is estimated to have a net count rate of 
 $(1.07\pm0.01)\times10^{-2}$ cts s$^{-1}$ in a circle of 18 arcsec radius centered 
 at the pulsar position.  The signal-to-noise of this elongated feature is 
 $\sim 4$ in the energy range $0.25-5$ keV. 

\begin{figure}
 \psfig{figure=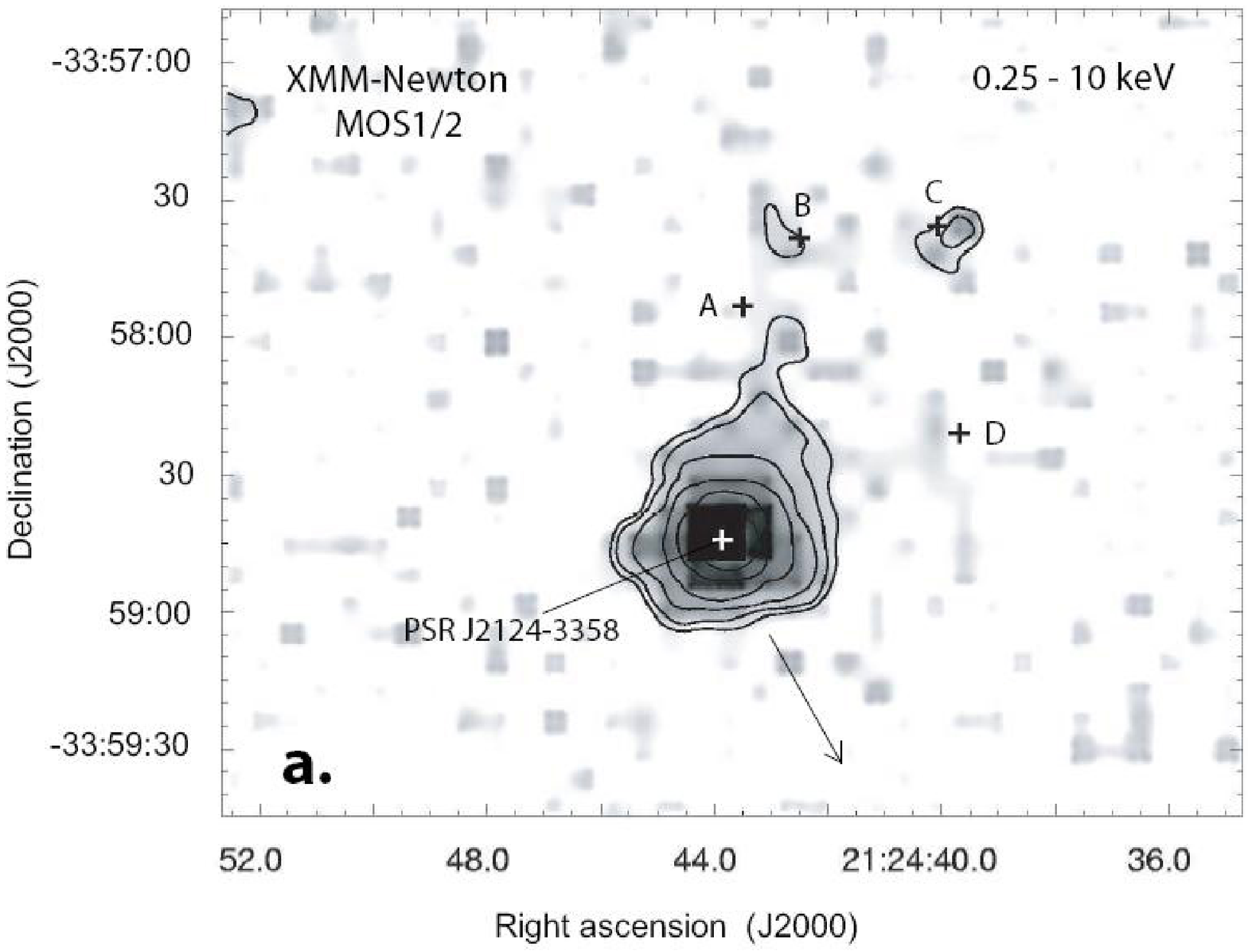,width=9.53cm,clip=}
 \psfig{figure=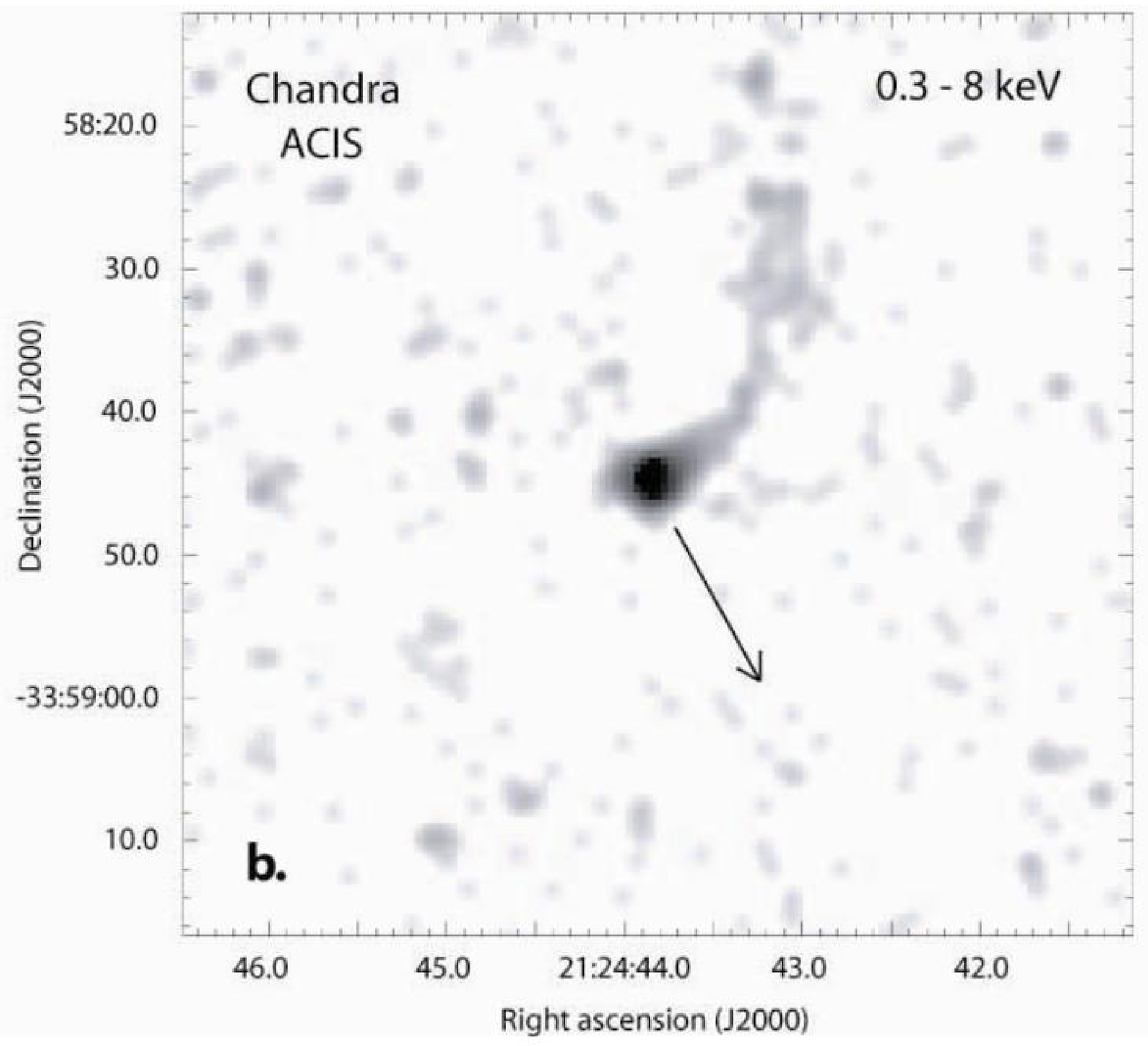,width=9.53cm,clip=}
 \caption{a) XMM-Newton MOS1/2 image of \PSR\ with overlaid contours. The pulsar proper 
  motion is indicated by an arrow.  The position of bright stars located in the 1.5 arcmin
  neighborhood of \PSR\ are indicated.  b) \PSR\ as seen by the ACIS detector aboard Chandra.}  
\end{figure}

 \PSR\ was  observed by Chandra with the back-illuminated ACIS-S3 chip in the focus of the XRT. 
 The observation took place on 2004 December 19-20 for an exposure time of $\sim 30$ ksec. An 
 image made from this data with 0.5 arcsec binning and adaptive smoothing applied (using a 
 Gaussian kernel with $\sigma < 1.5$ pixel) is shown in Figure 1b.  Arc-like diffuse emission 
 which is within the pulsar's H$_\alpha$ nebula is clearly detected. Its net count rate is
 estimated to be $(1.28\pm0.23)\times10^{-3}$ cts s$^{-1}$ in a box of $10\times 20$ arcsec 
 centered at RA(J2000)=$21^{\rm h} 24^{\rm m} 43.17^{\rm s}$, Dec=$-33^\circ 58' 33.53"$. With the aid of 
 PIMMS, the pileup fraction of the ACIS-S3 data was found to be less than $0.2\%$, i.e.~is 
 negligible.  The signal-to-noise of this feature is $\sim 5$ in the energy range of  
 $0.3-8$ keV.

 Since the PSF width of XMM-Newton is about 10 times that of Chandra, it blurred most of the 
 detailed structure seen in the Chandra data. However, it should be noted that the 
 overall direction of the feature in the Chandra image is consistent with the 
 orientation of the trail detected by XMM-Newton.

 In order to investigate a possible contribution to the diffuse X-ray emission by nearby 
 stars we investigated the Digitized Sky Survey data (DSS) for the sky region around 
 \PSR. There are four field stars (labeled as A, B, C and D in Figure 1a and Table 2) 
 in the 1.5 arcmin neighborhood of \PSR. From the USNO catalogue, we identified the 
 magnitudes of these stars. None of them is found to match the position of the diffuse
 elongated X-ray structure seen in the  XMM-Newton and Chandra data. It can be seen in 
 Figure 1a that the positions of  stars B and C coincide with two faint X-ray sources 
 which are disconnected with the trail emission of \PSR\, though. 

 \begin{table}
 \caption{Identifications of the stars around the X-ray trail of \PSR.}
 \begin{tabular}{lccc}
 \hline\hline
 Stars $^{a}$ & USNO catalogue ID & Magnitude: $R$ & Magnitude: $B$\\
 \hline
 A & U052543254607 & 17.1 & 18.7\\
 B & U052543254423 & 18.0 & 20.2\\
 C & U052543253997 & 16.7 & 17.4\\
 D & U052543253910 & 13.5 & 15.5\\
 \hline
 \end{tabular}
 a.~See Figure 1a.
 \end{table}

 To examine the nature of the faint X-ray nebular emission by a spectral analysis we make 
 use of the XMM-Newton data and extract the energy spectrum from within a $30\times35$ 
 arcsec box. Using a box rather than a circular selection region allows to avoid the 
 emission from the pulsar and excludes any potential contamination from the field 
 stars B and C. However, we estimate that the wing of the XMM-Newton PSF centered at 
 the pulsar position still contributes $\sim 20\%$ to the total counts inside the box. 
 The background  spectrum was extracted from a source free region near to \PSR. In total, 
 92 and 67 source counts are available from the trail in the MOS1 and MOS2 detectors, 
 respectively. Response files were computed by using the XMMSAS tasks RMFGEN and ARFGEN. 
 The spectra were dynamically binned so as to have at least 10 counts per bin. 

 In the Chandra data we selected the energy spectrum of the diffuse emission from a box 
 of size $10\times 20$ arcsec. Owing to the narrow PSF of Chandra the contamination 
 of pulsar emission in this box is negligible. The background spectrum was extracted 
 from a low count region near to the diffuse feature. In total 46 source counts are 
 contributed from the Chandra data. Response files were computed by using tools MKRMF 
 and MKARF of CIAO. The spectrum was binned to have at least 8 counts per bin. 
 The degradation of quantum efficiency of ACIS was corrected by applying
 the XSPEC model ACISABS.

 We hypothesize that the diffused emission should be originated from the interaction 
 of pulsar wind and the ISM. Synchrotron radiation from the ultra-relativistic electrons 
 is generally believed to be the emission mechanism of the pulsar wind nebula, which then
 is characterized by a power-law spectrum. With a view to test this hypothesis, we fitted 
 an absorbed power-law model to the nebular spectrum with XSPEC 11.3.1 in the $0.25-10$ keV
 energy range. With a column density of $5\times10^{20}$ cm$^{-2}$ as obtained from spectral 
 fits to the pulsar emission, we found that the model describes the observed spectrum fairly 
 well ($\chi^{2}_{\nu}=0.79$ for 26 D.O.F.). The photon index is $\alpha=2.2\pm0.4$ and the 
 normalization at 1 keV is $(2.94\pm0.48)\times10^{-6}$ photons keV$^{-1}$ cm$^{-2}$ s$^{-1}$ 
 ($1-\sigma$ error for 1 parameter in interest). In view of the low  photon statistic we tested 
 for a possible dependence of the fitted model parameters against the background spectrum. 
 All deviations found in repeating the fits with different background spectra were within the 
 $1-\sigma$ confidence interval quoted above. The unabsorbed fluxes and luminosities deduced 
 for the best fit model parameters and the energy ranges 0.1$-$2.4 keV and 0.5$-$10 keV are 
 $f_{X}=1.8\times10^{-14}$ ergs s$^{-1}$ cm$^{-2}$, $L_{X}=1.3\times10^{29}$ ergs s$^{-1}$ 
 and $f_{X}=1.2\times10^{-14}$ ergs s$^{-1}$  cm$^{-2}$, $L_{X}=8.9\times10^{28}$ ergs s$^{-1}$, 
 respectively. The best fitting spectral model is displayed in Figure 2. 

 \begin{figure}
 \centerline{\psfig{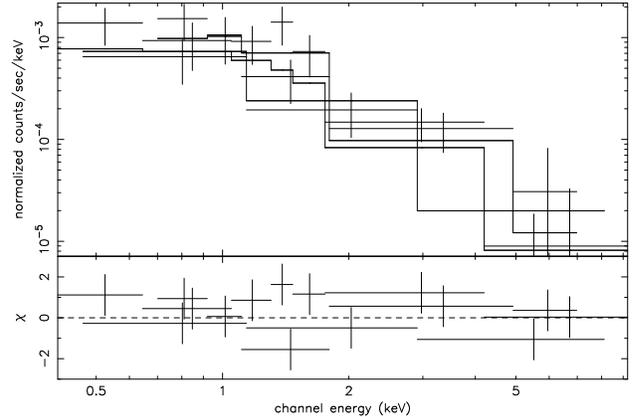}}
 \caption{Energy spectrum of the X-ray trail of \PSR\, as observed with
 MOS1/2 and ACIS-S3 detectors and simultaneously fitted to an absorbed power-law model
 ({\it upper panel}) and contribution to the $\chi^{2}$ fit statistic
 ({\it lower panel}).}
     \vspace{-0.4cm}
 \end{figure}

 Motivated by the extended source structure seen in \PSR\, we have also inspected the XMM-Newton 
 data of PSRs J0030$+$0451, J1024$-$0719 and J0437-4715 in order to search for extended tail-like 
 X-ray emission.  There is no evidence for any extended emission around these pulsars. In order 
 to deduce an upper limit for an X-ray trail in these sources we have estimated the net count 
 rates for the regions of $30\times35$ arcsec right behind the corresponding directions of 
 each pulsar's proper motion. The net count rate for any diffuse emission in these regions 
 is $(3.1\pm1.5)\times10^{-4}$ cts s$^{-1}$, $(3.1\pm0.7)\times10^{-4}$ cts s$^{-1}$ 
 and $(2.5\pm0.9)\times10^{-4}$ cts s$^{-1}$ for PSRs J0030$+$0451, J1024$-$0719 and J0437-4715, 
 respectively. The quoted errors are $1-\sigma$ confidence intervals. Assuming a Crab-like 
 spectrum (i.e.~$\alpha=2$) and taking the values of column density inferred from the pulsar 
 spectral fits ($1\times10^{20}$ cm$^{-2}$ for J0030$+$0451, $2\times10^{20}$ cm$^{-2}$ 
 for J1024$-$0719 and $3\times10^{19}$ cm$^{-2}$ for J0437-4715), we calculated the $3-\sigma$ 
 limiting fluxes from the above count rates. For PSRs J0030$+$0451, J1024$-$0719
 and J0437-4715, the limiting fluxes in 0.5$-$10 keV are estimated to be 
 $6.9\times10^{-15}$ ergs s$^{-1}$ cm$^{-2}$, $4.9\times10^{-15}$ ergs s$^{-1}$ cm$^{-2}$ 
 and $4.2\times10^{-15}$ ergs s$^{-1}$ cm$^{-2}$, respectively.

\section{Discussion \& Conclusion}

 Basic characteristics of the X-ray bow shocks detected around rotation-powered pulsars
 are summarized in Table 3. \PSR\ is a new candidate to this group. Its tail is found to 
 have an extent of $\sim 0.5$ arcmin. Adopting the distance of $\sim$ 250 pc, the tail
 has a length of $l\sim1.1 \times10^{17}$ cm. For the pulsar's proper motion velocity 
 of 58 km s$^{-1}$ (Manchester et al.~2005), the timescale, $t_{\rm flow}$, for the passage of 
 the pulsar over the length of its X-ray trail is estimated to be $\sim600$ yrs.  According 
 to the discussion in Becker et al.~(2005) on the trail emission of PSR B1929+10 we estimate
 the magnetic field in the shocked region by assuming $t_{\rm flow}$ to be comparable to the electron lifetime 
 of the synchrotron emission. This yields $\sim30\mu$G for the inferred magnetic field strength in the  
 emitting region. The magnetic field strength in the ISM is estimated to be $\sim2-6\mu$G (cf.~Beck et al.~2003 
 and references therein). Taking into account that the magnetic field in the termination shock might be 
 compressed (e.g.~Kennel \& Coroniti 1984), our estimation is approximately consistent if the compression 
 factor is $\sim 7$.

 \begin{table*}
 \begin{center}
 \caption{Properties of rotation-powered pulsars with X-ray/$H_\alpha$ bow shocks.}
 \begin{tabular}{ccccccccc}
 \hline\hline
 PSR & Other Name/ & \.{E} & $L_{X}^{diff}$ (0.5$-$10 keV)$^{a}$ & $\tau_{spin-down}$ & $H_\alpha$ & X-ray & Radio & Reference\\
     & associated SNR &erg s$^{-1}$ &erg s$^{-1}$ & years & bow shock & trail & nebula & \\
     \hline
     \multicolumn{9}{c}{Young pulsars }\\
     \hline
     J0537-6910 & LMC: N157B &$4.8\times10^{38}$ & $3.5\times10^{36}$  & $5.0\times10^{3}$  & - & d & - & 1,2  \\
     B1757-24   & Duck &$2.6\times10^{36}$ & $4.6\times10^{32}$ &  $1.6\times10^{4}$  & - & d & d & 3        \\
     B1853+01   & W44 &$4.3\times10^{35}$ & $5.7\times10^{32}$ & $2.0\times10^{4}$  & - & d & d & 4        \\
     J1747-2958 & Mouse &$2.5\times10^{36}$ & $5.3\times10^{34}$ & $2.6\times10^{4}$   & - & d & d & 5        \\
     B1951+32   & CTB80 &$3.7\times10^{36}$ & $2.5\times10^{33}$ & $1.1\times10^{5}$   & d   & d & d & 6,7,8      \\
     \hline
     \multicolumn{9}{c}{Middle age and old pulsars}\\
     \hline
     B0740-28   & &$1.4\times10^{35}$ & -  &  $1.6\times10^{5}$  & d & - & - & 9 \\
     B0633+17   & Geminga &$3.2\times10^{34}$ & $8.9\times10^{28}$ &  $3.4\times10^{5}$ & - & d & - & 10       \\
     B2224+65 & Guitar &$1.2\times10^{33}$ & $3.7\times10^{31}$  & $1.1\times10^{6}$  & d & ? & - & 11,12            \\
     B1929+10   &  &$3.9\times10^{33}$ & $8.3\times10^{29}$ &  $3.1\times10^{6}$     & - & d & ? & 13        \\
     \hline
     \multicolumn{9}{c}{Millisecond pulsars}\\
     \hline
     B1957+20   &  &$1.1\times10^{35}$ & $1.9\times10^{31}$ &  $2.2\times10^{9}$   & d & d & - & 14       \\
     J0437-4715 & &$3.9\times10^{33}$ & -  & $4.9\times10^{9}$  & d & - & - & 15,16 \\
     J2124-3358 &  &$4.4\times10^{33}$ & $10^{29}$  &  $5.9\times10^{9}$  & d & d & - & 17,18     \\
     \hline
     \end{tabular}
     \\
     \end{center}
     {\bf a.} The X-ray luminosities of the diffuse emission are taken from the corresponding reference and recalculate into energy band of 0.5$-$10 keV 
     for easy comparison.
     References: (1) Wang \& Gotthelf 1998; (2) Wang et al. 2001; (3) Kaspi et al. 2001; (4) Petre et al. 2002; (5) Gaensler et al. 2004; 
     (6) Migliazzo et al. 2002; (7) Moon et al. 2004; (8) Li et al. 2005; (9) Jones et al. 2002; (10) Caraveo et al. 2003; 
     (11) Romani et al. 1997; (12) Chatterjee \& Cordes 2002; (13) Becker et al. 2005; (14) Stappers et al. 2003; (15) Bell et al. 1995; 
     (16) Zavlin et al. 2002; (17) Gaensler et al. 2002; (18) This letter; and Manchester et al. 2005 otherwise.
     \end{table*}

 Following Becker et al.~(2005), we apply a simple one zone model (Chevalier 2000) to estimate the spectral 
 behavior and the X-ray luminosity of the nebular emission.  The X-ray luminosity and spectral index depend 
 on the inequality between the characteristic observed frequency $\nu^{\rm obs}_{X}$ and the electron synchrotron 
 cooling frequency $\nu_{\rm c}$ which is estimated to be $1.6\times10^{17}$ Hz. Since in general 
 $\nu^{\rm obs}_{X}>\nu_{\rm c}$, we concluded that the emission is in a fast cooling regime. Electrons with the 
 energy distribution, $N(\gamma)\propto\gamma^{-p}$, are able to radiate their energy in the trail with 
 photon index $\alpha=(p+2)/2$. The index $p$ due to shock acceleration typically lies between 2 and 3 
 (cf. Cheng, Taam, \& Wang 2004 and references therein). Taking $p=2.35$ yields $\alpha^{\rm th}\simeq2.2$ 
 which is in accordance with the result from the observed value $\alpha^{\rm obs}=2.2\pm0.4$. Assuming the 
 energy equipartition between the electron and proton (Cheng, Taam, \& Wang 2004), we take the fractional 
 energy density of electron $\epsilon_{\rm e}$ to be $\sim0.5$ and the fractional energy density of the magnetic
 field $\epsilon_{B}$ to be $\sim0.01$. Assuming a number density of ISM to be 1 cm$^{-3}$, the distance of 
 the shock from the pulsar is estimated to be $\sim3.6\times10^{16}$ cm. With these estimates, the calculated 
 luminosity, $\nu L_{\nu}$, is given as $\sim10^{29}$ ergs s$^{-1}$ which is well consistent with the observed 
 values of $1.3\times10^{29}$ ergs s$^{-1}$ (0.1-2.4 keV) and $8.9\times10^{28}$ ergs s$^{-1}$ (0.5-10 keV).

 Although the general properties of the X-ray trail in \PSR\, are not in contradiction with properties
 observed in other pulsars there are still ambiguities which are not completely resolved. 
 First, one should notice that the trail is misaligned with the direction of the pulsar's proper 
 motion. As reported by Gaensler, Jones, \& Stappers (2002), the head of the $H_\alpha$ bow shock is found 
 to be highly asymmetric about the pulsar's velocity vector with the apparent nebular symmetry axis deviated 
 from the velocity vector by $\sim30^{\circ}$. Even though the misalignment of the X-ray trail seems 
 to agree with the asymmetry of the $H_\alpha$ nebula, deeper observations by XMM-Newton and Chandra 
 are required in order to constrain the physical properties of this interesting nebula in higher detail.

\begin{acknowledgements}
 We thank the referee Giancarlo Cusumano for thoroughly reading the manuscript 
 and the many useful comments. We acknowledge the use of data obtained with XMM-Newton 
 and Chandra. XMM-Newton is an ESA science mission with instruments and 
 contributions directly funded by ESA Member States and NASA. 
\end{acknowledgements}


\begin{thebibliography}{}
\bibitem[]{} 
Bailes, M., Johnston, S., Bell, J. F., et al. 1997, ApJ, 481, 386
\bibitem[]{} 
Beck, R., Shukurov, A., Sokoloff, D., \& Wielebinski, R. 2003, A\&A, 411, 99
\bibitem[]{} 
Becker, W., \& Tr\"{u}mper, J. 1993, Nature, 365, 528
\bibitem[]{} 
Becker, W., \& Tr\"{u}mper, J. 1998, IAU Circular 6829
\bibitem[]{} 
Becker, W., \& Tr\"{u}mper, J. 1999, A\&A, 341, 803
\bibitem[]{} 
Becker, W., Tr\"{u}mper, J., Lommen, A. N., \& Backer, D. C. 2000, ApJ, 545, 1015
\bibitem[]{} 
Becker, W., Kramer, M., Jessner, A., et al. 2005, ApJ, in press (available from astro-ph/0506545)
\bibitem[]{}
Bell, J.F., Bailes, M., Manchester, R. N., Weisberg, J. M., \& Lyne, A. G. 1995, ApJ, 440, L81
\bibitem[]{} 
Caraveo, P. A., Bignami, G. F., DeLuca, A., Mereghetti, S., Pellizzoni, A., Mignani, R., Tur, A., \& Becker, W. 
2003, Science, 301, 1345
\bibitem[]{}
Chatterjee, S., \& Cordes, J.M. 2002, ApJ, 575, 407
\bibitem[]{} 
Cheng, K. S., Taam, R. E., \& Wang, W. 2004, ApJ, 617, 480
\bibitem[]{} 
Chevalier, R. A. 2000, ApJ, 539, L45
\bibitem[]{} 
D'Amico 2000, N., in Pulsar Astronomy - 2000 and Beyond, 202th ASP Conf. Ser., ed. M. Kramer, N. Wex \& R. Wielebinski (San Francisco:ASP), 27
\bibitem[]{} 
Gaensler, B. M., van der Swaluw, E., Camilo, F., Kaspi, V. M., Baganoff, F. K., Yusef-Zadeh, F., 
\& Manchester, R. N. 2004, ApJ, 616, 383
\bibitem[]{} 
Gaensler, B. M., Jones, D. H., \& Stappers, B. W. 2002, ApJ, 580, L137
\bibitem[]{}
Johnston, S., Lorimer, D. R., Harrison, P. A., et al. 1993, Nature, 361, 613
\bibitem[]{}
Jones, D.H., Stappers, B.W., \& Gaensler, B.M. 2002, A\&A, 389, L1
\bibitem[]{} 
Kaspi, V. M., Gotthelf, E. V., Gaensler, B. M., \& Lyutikov, M. 2001, ApJ, 562, L163
\bibitem[]{} 
Kennel, C. F., \& Coroniti, F. V. 1984, ApJ, 283, 694
\bibitem[]{} 
Li, X. H., Lu, F. J., \& Li, T. P. 2005, ApJ, 628, 931
\bibitem[]{} 
Manchester, R. N., Hobbs, G. B., Teoh, A., \& Hobbs, M. 2005, AJ, 129, 1993 
\bibitem[]{} 
Migliazzo, J. M., Gaensler, B. M., Backer, D. C., Stappers, B. W., van der Swaluw, E., \& Strom, R. G. 2002, ApJ, 567, L141
\bibitem[]{} 
Moon, D.-S., Lee, J.-J., Eikenberry, S. S., et al. 2004, ApJ, 610, L33
\bibitem[]{} 
Petre, R., Kuntz, K. D., \& Shelton, R. L. 2002, ApJ, 579, 404
\bibitem[]{}
Romani, R.W., Cordes, J.M., \& Yadigaroglu, I.-A. 1997, ApJ, 484, L137
\bibitem[]{} 
Somer, A. 2000, in Pulsar Astronomy - 2000 and Beyond, 202th ASP Conf. Ser., ed. M. Kramer, N. Wex \& R. Wielebinski (San Francisco:ASP), 17
\bibitem[]{} 
Stappers, B. W., Gaensler, B. M., Kaspi, V. M., van der Klis, M., \& Lewin, W. H. G. 2003, Science, 299, 1372
\bibitem[]{}
Wang, Q.D., \& Gotthelf, E.V. 1998, ApJ, 494, 623
\bibitem[]{}
Wang, Q.D., Gotthelf, E.V., Chu, Y.-H., \& Dickel, J.R. 2001, ApJ, 559, 275
\bibitem[]{}
Zavlin, V.E., Pavlov, G.G., Sanwal, D., Manchester, R.N., Tr\"{u}mper, J., Halpern, J.P., \& Becker, W. 2002, ApJ, 569, 894
\end{thebibliography}
\end{document}